\documentclass{vldb}

\usepackage{amsmath}
\usepackage{amssymb}
\usepackage{times}
\usepackage{boxedminipage}
\usepackage{xspace}
\usepackage{array}
\usepackage{epsfig}
\usepackage{calc}
\usepackage{multirow}
\usepackage{rotating}
\usepackage{enumitem}
\usepackage[usenames,dvipsnames]{color}
\usepackage{tabularx}
\usepackage{balance}  
\usepackage{tikz}
\usepackage{wasysym}
\usepackage[hyphenbreaks]{breakurl}
\usepackage{xcolor}
\usepackage[font={small}]{caption}
\usepackage{alltt}
\usepackage{setspace}
\usepackage{wrapfig}
\newcommand{\subparagraph}{} 
\usepackage{etoolbox}

\definecolor{commentgray}{rgb}{0.2,0.2,0.2}
\definecolor{bggray}{rgb}{0.98,0.98,0.98}
\usepackage{listings}
\usepackage[final]{microtype}
\usepackage[scaled]{inconsolata}
\usepackage[T1]{fontenc}
\usepackage{graphicx}
\usepackage{subfig}
\usepackage{soul}
\usepackage{pifont}

\usepackage[
            bookmarksopen=true, 
            pdfhighlight=/I,
            pdfpagemode=UseOutlines, 
            linkcolor=blue, 
            pdfborder={ 0 0 0 },
            pageanchor=false]{hyperref}
\hypersetup{
    pdfauthor = { },
    pdftitle = {},
    pdfkeywords = { },
    pdfborder={ 0 0 0 }
}

\usepackage{float}
\usepackage{bookmark}
\usepackage{amsmath,amsfonts,amssymb,stmaryrd}
\usepackage{latexsym}
\usepackage{multirow}
\usepackage{emptypage}
\usepackage{booktabs}
\usepackage{enumerate}
\usepackage{braket}
\usepackage{algorithm}

\usepackage[capitalize,nameinlink]{cleveref}

\Crefname{section}{Section}{Sections}
\Crefname{figure}{Figure}{Figures}

\usetikzlibrary{
  decorations.pathreplacing,
  angles,
  quotes,
  arrows,
  positioning,
  shapes.symbols,
  shapes.callouts,
  patterns,
  fit
}

\newcommand{\wiser}{WiSer\xspace}

\robustify{\cite}

\newcommand{\squishitemize}{
 \begin{list}{$\bullet$}
  { \setlength{\itemsep}{0pt}
     \setlength{\parsep}{3pt}
     \setlength{\topsep}{3pt}
     \setlength{\partopsep}{0pt}
     \setlength{\leftmargin}{1.95em}
     \setlength{\labelwidth}{1.5em}
     \setlength{\labelsep}{0.5em} } }

\newcounter{Lcount}
\newcommand{\squishlist}{
    \begin{list}{\arabic{Lcount}. }
   { \usecounter{Lcount}
        \setlength{\itemsep}{0pt}
        \setlength{\parsep}{3pt}
        \setlength{\topsep}{3pt}
        \setlength{\partopsep}{0pt}
        \setlength{\leftmargin}{2em}
        \setlength{\labelwidth}{1.5em}
        \setlength{\labelsep}{0.5em} } }

\newcommand{\squishend}{\end{list}}

\definecolor{todo-color}{rgb}{1,0,0}

\definecolor{yingjun-color}{rgb}{0,0,1}

\definecolor{comment-color}{rgb}{0.25,0.25,0.25}

\definecolor{my-blue}{RGB}{79,117,173}
\definecolor{my-green}{RGB}{93,166,108}
\definecolor{my-red}{RGB}{191,79,85}
\definecolor{my-red-2}{RGB}{187,28,43}
\definecolor{my-salmon}{RGB}{241,121,139}
\definecolor{my-purple}{RGB}{128,117,175}
\definecolor{bgblue}{RGB}{209,235,254}
\definecolor{bggreen}{RGB}{209,232,221}
\definecolor{bgred}{RGB}{254,212,210}

\begin{document}

\newcommand{\mail}[1]{\href{mailto:#1}{#1}}
\renewcommand{\thefootnote}{\fnsymbol{footnote}}

\title{WiSer: A Highly Available HTAP DBMS 
for IoT Applications}
\author{
Ronald Barber$^1$, Christian Garcia-Arellano$^2$, Ronen Grosman$^2$, Guy Lohman$^1$,\\ 
C. Mohan$^1$,  Rene Muller$^4$, Hamid Pirahesh$^1$, Vijayshankar Raman$^1$,\\
Richard Sidle$^1$, Adam Storm$^2$, Yuanyuan Tian$^1$, Pinar Tozun$^3$, Yingjun Wu$^1$\\
\small {\em  $^1$IBM Almaden Research Center \quad
          $^2$IBM Analytics}\\
\small {\em  $^3$IT University of Copenhagen \quad
          $^4$Bern University of Applied Sciences BFH} \\ [2mm]
}

\maketitle

\begin{abstract}

In a classic transactional distributed database management system (DBMS), 
write transactions invariably synchronize with a coordinator 
before final commitment.
While enforcing serializability, this model has long been criticized 
for not satisfying the applications' availability requirements.
When entering the era of Internet of Things (IoT), this problem has become 
more severe, 
as an increasing number of applications call for the capability of 
hybrid transactional
and analytical processing (HTAP), 
where aggregation constraints need to be enforced as part of transactions.
Current systems work around this by creating escrows, 
allowing occasional overshoots of constraints, 
which are handled via compensating application logic.

The \wiser DBMS targets consistency with availability,
by splitting the database commit into two steps.
First, a \textsc{Promise} step that corresponds to what humans are used to as commitment, and runs
without talking to a coordinator. Second, 
a \textsc{Serialize} step, that fixes 
transactions' positions in the serializable order, via a consensus procedure. 
We achieve this split via 
a novel data representation that embeds read-sets into transaction deltas, 
and serialization sequence numbers into table rows.
\wiser does no sharding (all nodes can run transactions that modify
the entire database), and yet enforces aggregation constraints.
Both read-write conflicts
and aggregation constraint violations are resolved lazily in the serialized data.
\wiser also covers node joins and departures as database tables, 
thus simplifying correctness and failure handling. 
We present the design of \wiser as well as experiments suggesting this
approach has promise.

\end{abstract}

\maketitle

\section{Introduction}
\label{introduction}

For decades, database management systems (DBMSs) have separated on-line transaction
processing (OLTP) from on-line analytics processing (OLAP) in distinct systems.
This has been more an engineering compromise than a desirable feature: 
OLTP systems typically modified small portions of the database and demanded low latency,
whereas OLAP systems typically queried large swaths of the database in read-only mode and 
demanded high throughput. But the promise of
classical DBMS transaction theory and its serializability concepts is that \textit{all} transactions --
even with arbitrarily complex queries embedded within them -- can be handled correctly (i.e., serializably) by a single DBMS, so that applications can program against a single-system image, even while running
in a distributed system with high concurrency.

Increasingly over the last few years, the Internet-of-Things (IoT) applications are gaining their popularity. 
They have begun to demand this ideal of 
doing analytics on the same system as transaction processing, and even
doing analytics within the transactions themselves ~\cite{gartnerHTAP,ozcan2017hybrid}.
This is known as hybrid transaction and analytical processing (HTAP).
However, even with significant advances in the speed and capacity of the underlying hardware, 
true HTAP in a distributed environment is extremely challenging. 
If a transaction, running on any node, can
query the entire database state up to that point in time (in
serializable order) and subsequently make a modification that might touch
the entire database, then that transaction will effectively need to lock the entire database while it runs. Not only can such transactions run for a long time, but their locking behavior excludes other transactions, 
killing concurrency and slowing performance to a crawl. 

Database sharding across distributed nodes somewhat eases the problem by partitioning the database into
shards and similarly partitioning the transaction workload, so that most transactions are local
to a single shard. But often aggregation constraints spanning multiple nodes need to be enforced, and 
this usually forces escrowing transactions while aggregating across shards.

Let us consider an online shopping application that is at the core of 
the TPC-C and TPC-E transactional benchmarks and captures the full HTAP challenge. 
It has two tables:

$\bullet$ \texttt{Products(productId, unitPrice)}: containing real goods (as in TPC-C),
securities (as in TPC-E), or services (e.g.,  a slot on a delivery truck).

$\bullet$ \texttt{Orders(orderId, productId, qty, price)}: containing orders for a given quantity
of a given product, at a given price. Customer orders express quantity as a
\textit{negative} number, while restocking orders have \textit{positive} values for quantity.

Now consider we want to run the following two transactions on these tables:

$\bullet$ \texttt{NewOrder}: Look up current prices for a few products, and
insert rows into \texttt{Orders} table with a certain order ID.

$\bullet$ \texttt{UpdatePrice}: Change the unit price for some products
in the \texttt{Products} table. 

Note that, unlike TPC-C, we do not have an ``Inventory'' table holding 
a materialized running overall quantity of product; instead, the transactions 
need to check an \textit{aggregation constraint} ensuring that the sum of
all quantities cannot be negative for any product.
The lack of ``Inventory'' table is common in modern IoT applications. 
As \texttt{NewOrder} can come from edge devices (e.g., mobile phones) from across the world, 
maintaining such an ``Inventory'' table could easily lead to centralized contention, 
caused by both 2-phase locking (2PL) and 2-phase commit (2PC).
In such an application, it is impossible to escrow the aggregation constraints;
In addition, it is also not realistic to shard the database, as 
these aggregation constraints are not ``shardable''. 


Instead of maintaining aggregation views in the DBMS, these constraints are invariably delegated to 
application logic. This brings us to the limitations expressed by the well-known CAP theorem by Brewer, which states that it is impossible for a distributed data store to simultaneously provide more than two out of the following three guarantees: Consistency (C), Availability (A), and tolerance to Partitions of the network connecting nodes (P).

\subsection{CAP Theorem within DBMS}

Brewer observes that applications always prioritize availability, even
applications that modify complex distributed state with many concurrent
actors~\cite{brewerCAPa,brewerCAPb}. A transaction cannot always 
wait to access the full global state, and cannot synchronously modify 
this state in a way that is immediately reflected globally.
Instead, there is complex compensatory logic in the application to handle
constraint violations due to concurrent activity.

A transaction is a contract between two parties, e.g., 
between a customer and a
retailer. When the customer clicks the ``submit'' button, they are committing, not just in
the DBMS sense but also in the legal contract sense, to buy these products at
these prices; conversely, the retailer is committing to supply.

But, the {\em serializable order of these transactions is established after commit}.
In a retail application, this could be the sequence in which orders are filled
by consuming inventory or space on a delivery truck. Constraints
can and do get violated all the time, and are resolved by compensation, e.g., via
an apologetic email about a ``back-ordered'' item and perhaps a coupon.

Thus, commit always has ``asterisks'' (contingencies): the parties agree that the contract is
contingent on the global state (reflecting all concurrent transactions, in
serializable order) not differing too much from the state seen at commit. Banks
usually state this asterisk explicitly, e.g., by listing their algorithm for
processing overdrafts.
The problem with this application logic is that it complicates reasoning about database
atomicity or durability.

\subsection{HA and HTAP within DBMS}

The Wildfire-Serializable (\wiser) project seeks to provide both high availability (HA) and HTAP within the same distributed DBMS 
without sharding, thus extending atomicity and durability to the handling of consistency contingencies.  \wiser is a follow-on project to the Wildfire research project, which developed high-throughput HTAP at
snapshot isolation and has been commercialized as the 
IBM DB2 Event Store~\cite{barber2016wildfire,barber2017wildfire}.  
Now we first give an overview of the novel way in which \wiser commits a transaction
to achieve this significant improvement.

\subsubsection{Symbolic and Resolved Commit}

A transaction in \wiser can run on any node, even on poorly connected nodes. Commit is
split into two stages:  


$\bullet$ \textbf{\textsc{Promise}} stage, after which the deltas (i.e., state changes) of a transaction $X$
are persisted durably and the transaction is ``symbolically committed''. There
are three ``symbolic'' entities in the delta that are resolved lazily:  (1) the
position of $X$'s delta in the serializable order; (2) $X$'s conflict status -- whether
the queries in the transaction saw the latest state (per the serializable
order); and (3) $X$'s constraint status -- whether the database state up to $X$ satisfies
aggregation constraints. These symbols capture the contingency semantics: 
if $X$ confronts access conflicts or violates constraints,
$X$ will roll back and the
application can specify one or more compensatory actions. 

$\bullet$ \textbf{\textsc{Serialize}} stage, to pick the position of 
the $X$'s delta in the serializable
order, which is done via consensus. 
This stage inherits the HA and liveness
properties of that consensus protocol. 


\wiser has further asynchronous stages, to resolve
$X$'s conflict violation, to
resolve $X$'s constraint violation, and to ``publish'' $X$'s 
delta in a query-efficient form. 
But $X$'s status is deterministic after the \textsc{Serialize} stage. 
Thus, an application
running on a poorly connected node, as long as it can get to serialize, can
always run a query to find the transaction status. 

\subsubsection{HA HTAP via Consensus}

Conceptually, doing HTAP with HA is just a consensus problem. Imagine a
database that is just a huge Raft~\cite{ongaro2014search} log. Transactions append deltas to this
log. In addition to these changes to user tables, each delta must also capture 
additional metadata and cluster state in this log:


$\bullet$ Transaction snapshot and read-sets: the queries in a transaction $X$ are
run as of a snapshot, which is a prefix of this log. To resolve
$X$'s conflicts, we need to know this snapshot and the nature of $X$'s
queries to compute whether the query result would have changed due to deltas
between the snapshot that $X$ sees and the position of $X$'s delta. Often this ``read-set''
information is stored in custom data structures and lock tables, making it
harder to reason about correctness in failure cases.

$\bullet$ Node departures and (re-)joins: say a node $N$ ``hangs'' for a while, and so its
peers evict it. How do they do that, while atomically stopping that node from
adding deltas to the log? We need to add the ``node departure'' event to the log as well.



\begin{figure}[th]
\centering
\includegraphics[width=\columnwidth]{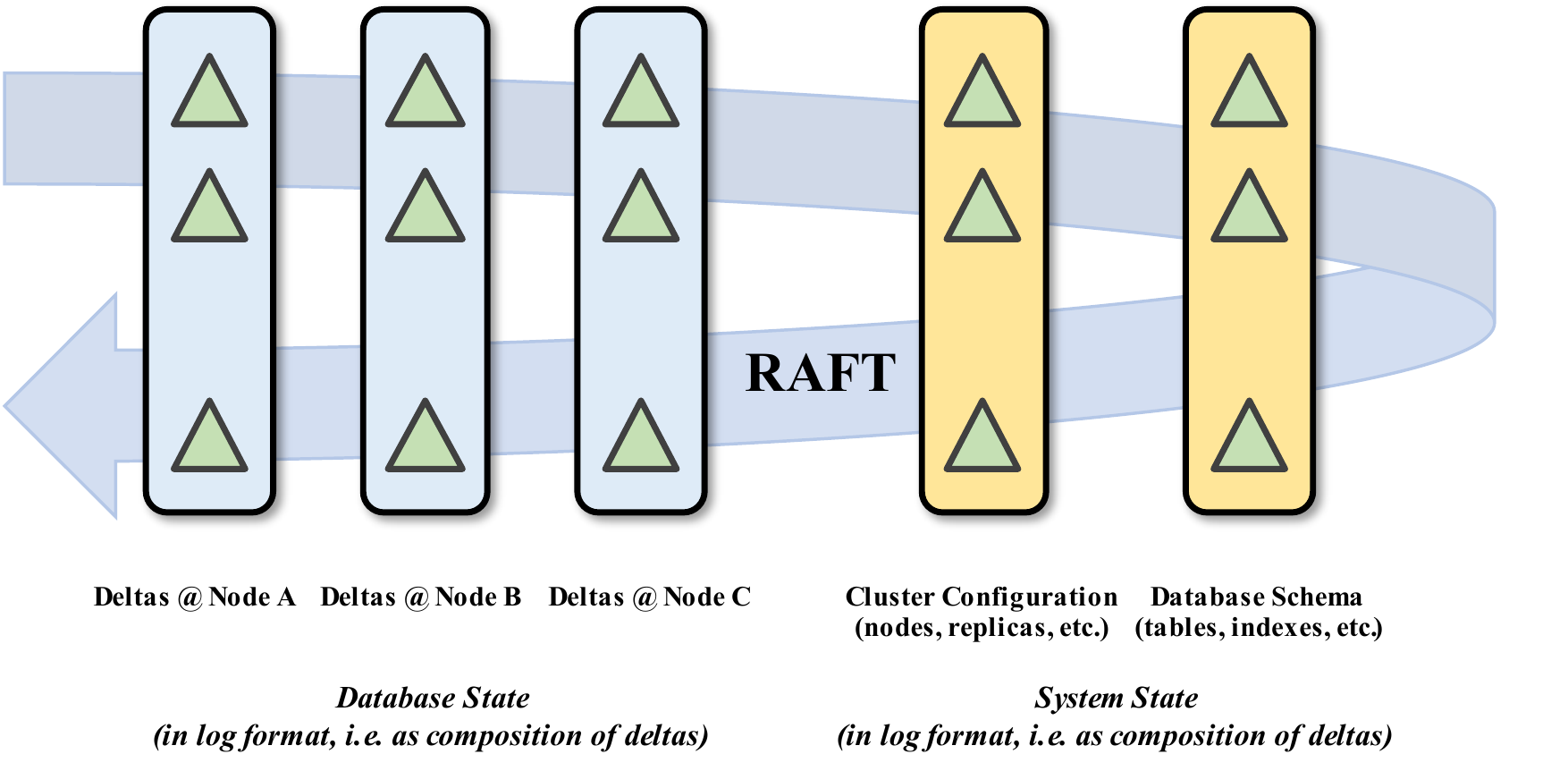}
\caption{If all state changes (database and system) are in a durable,
consistent, HA consensus log, we have a serializable system.}
\label{globalRaftLog}
\end{figure}

\cref{globalRaftLog} illustrates this data model. 
If we had such a consensus log,
then in theory we have HTAP and HA -- transactions will run serializably just by
querying the log, and they will make progress even in poorly connected
situations, following the HA behavior of the consensus protocol.
Of course, such a grand consensus log would be too slow, both in throughput and
latency. This in turn would cause too many rollbacks, because longer-running
transactions are more likely to have read-write conflicts.

\begin{figure}[th]
\centering
\includegraphics[width=0.9\columnwidth]{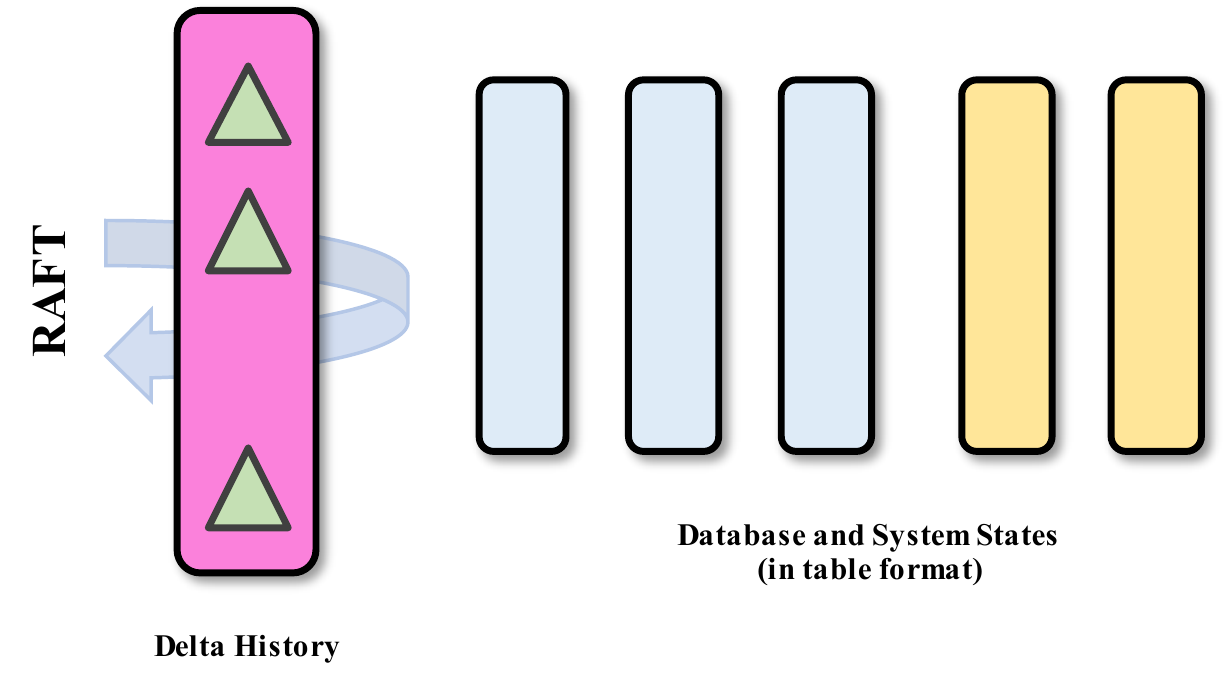}
\caption{\wiser puts only the history of state changes in a Raft log, 
and stores all other information in database tables.}
\label{globalRaftLogWiser}
\end{figure}

\wiser instead picks the other extreme. As \cref{globalRaftLogWiser} shows, 
\wiser has a consensus log that tracks only the history of the state changes. 
All the other state is in database tables, 
and our challenge is to get consensus on changes
to this other state using only the change log.

\subsubsection{High Throughput Scale-out}

\wiser's design of resolving conflicts and constraints after serialize also
gives a significant throughput benefit.

\textsc{Promise} is a local operation, with replication for durability alone, and does
no locking, so it scales out quite readily;
\textsc{Serialize} is a consensus operation, whose cost is independent of the
transaction load. We have an elected \textsc{Serializer} (or leader), 
who appends
the log-sequence number (LSN) ranges to a broadcasted file every $N$ time intervals,
where $N$ is chosen based on the desired latency. Each serialization picks the
serial order for all the new transactions, across all nodes, that are available to
\textsc{Serialize} at that time.

Thus, the key scalability challenge is in the conflict resolution step
performed after \textsc{Serialize}, to resolve the symbolic fields. 
Like Calvin~\cite{thomson2012calvin},
\wiser's database state is deterministic upon \textsc{Serialize}. 
So constraints and conflicts are resolved on a non-changing database state. 
This allows conflict
resolution to be done at high throughput, and to be trivially scalable.

\subsubsection{Efficiency of Aggregation Constraints}

In the previous online shopping application, 
suppose \texttt{NewOrder} transaction had to compute the inventory 
(via an aggregation query) and verify that it will remain positive 
after this order. The query will
take so long to run that the resulting transaction will surely conflict with
some \texttt{UpdatePrice} transaction that snuck through in the meantime. This is why TPC-C uses
materialized aggregation views, but of course they involve sharding.

\wiser instead uses lazy \textsc{ConstraintResolvers} to resolve constraints. 
They run the aggregation as a
query each time, so that \texttt{NewOrder} transaction is not burdened with having to update
inventories and can scale easily.
The \textsc{ConstraintResolvers} keep up with the transaction rate because they run as
streaming queries, continually streaming the serialized, conflict-resolved
deltas through streaming queries to verify aggregation constraints. 

\subsection{Paper outline}

\cref{sec:design} describes the design of \wiser, focusing just on getting to the serial order.
\cref{sec:conflicts} and \cref{sec:constraints} discuss how \wiser handles
conflict and constraint resolution. 
We describe the system status and experiment results in \cref{expts}.
\cref{sec:related-work} reviews related work and 
\cref{sec:conclusion} concludes the paper.

\section{Design}
\label{sec:design}

We model a database as a logical \textit{chain} of state modifications, called
\textit{deltas}. The chain is in time order, that is, smaller prefixes of the chain
correspond to earlier states of the database. 
These state modifications are made by programs called \textit{transactions} that
logically read the current database state (the chain prefix up to that modification),
and accordingly apply transaction logic to do the modification\footnote{
Application programs further need that transactions must be \textit{serialized} in an
order that respects the user time at which they were executed. We defer this
problem until \cref{subsub:strictser}.}. 
Each transaction runs local to a node, but may 
query the database state across nodes. 
We call the deltas produced at each node as a \textit{subchain} and the 
database chain is an interleaving of these subchains. 
\wiser performs background
replication and hardening of each subchain, as described later.

\begin{figure}[th]
\centering
\includegraphics[width=0.8\columnwidth]{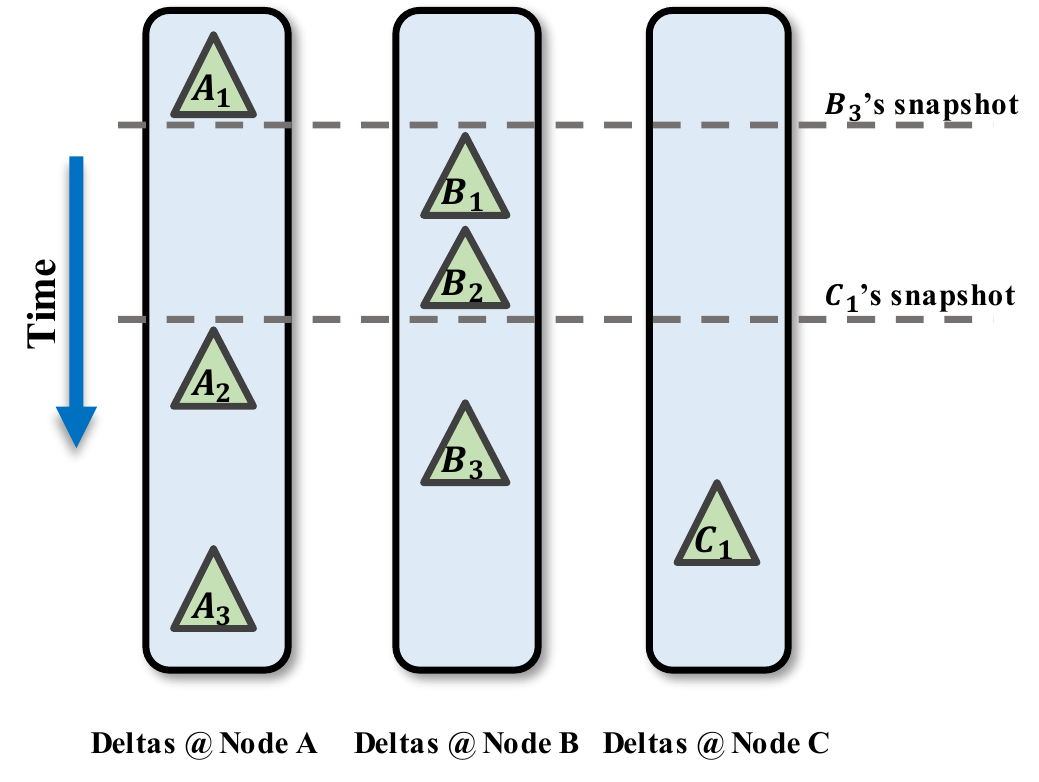}
\caption{
Transactions run at three nodes. A transaction only see 
deltas made before their snapshot.}
\label{chains}
\end{figure}

\cref{chains} shows an example with three nodes, 
$A$, $B$, $C$, 
and transactions' deltas produced at these nodes: 
$A_1, A_2, ..., B_1, ..., C_1, ...$. 
The database state is the composition of
deltas that come in time order: 
$A_1 \rightarrow B_1 \rightarrow B_2 \rightarrow A_2
\rightarrow B_3\rightarrow C_1 \rightarrow A_3
...$.

\textbf{Content of a Delta.}
\wiser allows only one kind of state modification: an \textit{upsert} of a row to a table.
An update-by-key is treated as a query to get the prior row content, followed by
an upsert of the modified contents. Deletes are updates that set an \texttt{isDeleted} field.
A transaction $X$ can run many queries, and do many (or zero) upserts.
The delta encapsulates not just these upserts, but also the
\textit{read-set} of $X$ -- the read queries the transaction $X$ ran.
Logically, the queries in
$X$ should read the database chain up to $X$; that is, $X$ shall see 
any updates that happen before it. But in practice, the queries in $X$ read a prefix of this chain,
called $X$'s \textit{snapshot}: as shown in \cref{chains}, $B_3$ 
can observe the delta made by $A_1$, but does not see that made by $A_2$, which happens after the snapshot.
If these query results can change due to intervening deltas (between $X$'s snapshot and $X$), $X$ 
must roll back. Therefore we add the read-set and 
the snapshot, as a serialization sequence number, to the delta.

\wiser represents transaction deltas in log format, as a Parquet~\cite{parquet} row-group 
per modified table, all concatenated together into a byte-string.
Read-sets are represented as upserts against a \texttt{ReadWriteSet} table, whose format 
we explain in \cref{sec:conflicts}.

\textbf{Consensus for Serial Order of Writes.}
\wiser uses Raft~\cite{ongaro2014search} for achieving consensus on this serial order. But we do
not directly put these deltas, or even the metadata, such as the range of log sequence numbers (LSNs) they span, onto a Raft
log. Instead, our Raft log only contains one entry each time a new \textsc{Serializer}
(or leader) is elected. This log holds a
\texttt{Serializers} table, with schema \texttt{(SerializerNodeID, SerializerSeqNum,
StartingBatch)}, which is initialized to a single row $(0, 0, 0)$ at database
creation (e.g., node $A$ is the first leader). Nodes are assumed to have unique
IDs. \texttt{SerializerSeqNum} is an increasing number $0, 1, 2, ...$
which is increased when a new \textsc{Serializer} is elected, and 
\texttt{StartingBatch} is an offset into a separate
\texttt{SerializeFrontiers} table.

\texttt{SerializeFrontiers} holds the serial order of deltas. 
Each row corresponds to a serialization batch, 
and contains the node ID
originating a batch, and the LSN range (only the upper bound)
of the deltas in that batch.
Nodes repeatedly run batches of transactions, harden their deltas,
and send the LSN range to the current \textsc{Serializer}. The 
\textsc{Serializer} picks the serial order and records it by appending to \texttt{SerializeFrontiers} table. 
By inspecting the tail of \texttt{SerializeFrontiers} table, 
a query can determine 
what is serialized across the whole database. 

\texttt{SerializeFrontiers} table is implemented as a list of files: 
a \textsc{Serializer}
with node ID $N$ and sequence number $S$ writes to file named $SerFront.N.S$ (this file is
lazily replicated to all nodes). Upon a failure and/or a network partition,
multiple nodes can think they are the \textsc{Serializer}, but the Raft log \texttt{Serializers}
picks the winner -- the row with maximum \texttt{SerializerSeqNum}, with ties broken by
picking the earliest entry. 

\textbf{Visibility and Conflict Rollbacks.}
Of course, transactions do not just do blind writes. They run queries (including both reads and writes)
as well. As explained in \cref{chains}, all queries within a transaction see a common prior state, called
its snapshot. This raises two challenges:

$\bullet$ Visibility: the transaction snapshot needs to be recent enough (not too
stale), highly available (when individual nodes are down), and often needs
auxiliary structures (indexes) to perform acceptably. 


$\bullet$ Read-write conflicts: after determining the serializable order (via
consensus), \wiser need to roll back all transactions whose reads have been changed in
value between the seen snapshot and the serialization point. As discussed above, 
\wiser append the read-set of each transaction to its delta. This
allows precise determination of read-write conflicts.

\cref{sec:conflicts} describes how we address both these challenges.
Note that we do not call out write-write conflicts (concurrent transactions
updating rows with the same key) separately. In \wiser, a write of a row into a
table with a defined primary key automatically performs a read of that row
(i.e., all writes are upserts), thus handling all conflicts as read-write.

\textbf{Constraint Rollbacks.}
State modifications must also respect aggregation constraints, such as the rule
in the online shopping application that inventory for each product must stay positive. We require
aggregation constraints to be algebraic (that is, involving sums, counts, and
averages, but not medians), so as to allow parallel as well as
incremental computation.  

Even with algebraic aggregates, constraint resolution is challenging because 
aggregates are generally not monotonic. For example, a batch of transactions may start with 
some orders that exceed the available inventory, and then a restocking, and then some
more orders. 
\cref{sec:constraints} describes how we tackle this challenge, and presents a parallel algorithm 
for incremental constraint resolution.


\subsection{Lazy Transaction Resolution}
\label{stages}

The design presented above involves a bunch of background tasks that happen in parallel with
transaction execution: replicating deltas, serializing them, resolving conflicts, and resolving constraints.
As these background tasks progress, they take a transaction through various stages, which we now detail.

\textbf{Stage 1: \textsc{Promise}.} A transaction submitted
by a client is first assigned to a subchain, and accordingly sent to a node.
Clients can give a hint as to where to run a transaction, that is, based
on what rows the transaction is inserting, to get a partitioning that queries can exploit.
But this is just a hint: every transaction can modify the entire database.

The transaction delta is appended to a node-local log, which is implemented as a 
a lock-free list of blocks.
\wiser asynchronously and automatically replicates this log to replica nodes. Promise completes
when a quorum of replicas has received and hardened the log, which is tracked via heartbeats.
\textsc{Promise} needs no global cluster communication; AP
applications stop at this stage. 

\textbf{Stage 2: \textsc{Serialize}.} Nodes continually inform an elected
\textsc{Serializer} of their log growth in the form of LSN 
ranges. The \textsc{Serializer} repeatedly picks a node and publishes its recent delta
to the global serial order, as discussed earlier. 
Other serial orders are also acceptable, such as ordering by the local commit timestamps.
This policy serializes entire
batches of changes from each node in a deterministic manner, similar to 
Calvin~\cite{thomson2012calvin}. 

The database state up to the serialization frontier is \textit{deterministic}.
Each delta has two symbolic fields:

$\bullet$ Conflict status: Transactions may roll back, 
due to read-write conflicts.
But since the read-set is embedded in the delta, this resolution is deterministic.
We use a table called \texttt{Rollbacks} (appended to by the
\textsc{ConflictResolver}), to cache this resolution.

$\bullet$ Constraint status: The
\textsc{ConstraintResolver} lazily checks aggregation constraints and appends to a
\texttt{ConstraintFailures} table.  Queries use these two tables,
which are essentially materialized views,
to efficiently view the latest serialized database state.

Applications requiring strict serializability can stop at \textsc{Serialize} stage. \wiser
does wait for publish and constraint resolution so that it can give a 
transaction status to the client, but this can time-out.
The semantics upon such a time-out is identical to that of
time-out at commit in a classical DBMS.

\textbf{Stage 3: \textsc{Conflict Resolve}.} A background \textsc{ConflictResolver} uses
the read-sets embedded in transaction deltas (across all nodes) 
to check if the value of any of these reads has changed by
the time the transaction was serialized. 
In our current implementation, the \textsc{Serializer} performs the
role of \textsc{ConflictResolver} as well.

\textbf{Stage 4: \textsc{Published}.} A background \textsc{Publisher}
runs at each node 
to make the serialized data efficiently accessible to point queries.

\textbf{Stage 5: \textsc{Constraint Resolve}.} 
A background \textsc{ConstraintResolver} continually reads 
the conflict-resolved deltas, across all nodes, to
verify aggregation constraints. It is at this point that
the transaction status can be returned to a client.

\textbf{Stage 6: \textsc{Groom to Cloud}.} Publisher writes out very small
blocks, because we need low latency. This is too expensive for OLAP, 
so we perform another round of data reorganization to consolidate blocks
and to migrate them to (less expensive) cloud storage.

\subsubsection{Strict Serializability}
\label{subsub:strictser}

Applications may expect more than serializability: 
if a transaction $T_1$ returns to the application at 10 am, 
a transaction $T_2$ issued at 10.01 am to a different node must see $T_1$'s writes. 
This is called strict serializability.
This stricter guarantee is readily available in \wiser, provided the application
waits until its original transaction reaches \textsc{Serialize} stage. 
By definition, any successful transaction (and query) 
submitted after that point must see all serialized changes, 
so will see the original transaction.
Of course, if the original transaction ran on a poorly connected node and the
application timed-out before \textsc{Serialize}, it has to keep polling to check if its
transaction did get serialized successfully.

\subsection{Visibility and Efficient Publish}
\label{sub:visibility}

Once a transaction has been serialized, it needs to be visible to any subsequent query.
Technically, the query can read \texttt{SerializeFrontiers} table, 
accordingly contact individual nodes for segments of their local log,
and thus scan the entire history of database deltas. This, with lookups (antijoins) into
\texttt{Rollbacks} and \texttt{ConstraintFailures} tables, gives full visibility to the serialized
state.

But OLTP applications have very strict latency requirements, 
and timeouts are perceived
as failures. Scanning the entire log is prohibitively slow.
This is why every node runs a \textsc{Publisher} that scans the recent appends to the log at
that node and converts them to a query-efficient form.
This involves three pieces:

$\bullet$ Organizing data by tables: \textsc{Publisher} separates the deltas by table
(recall that every modification is an upsert), and appends them to per-table 
blocks, still in Parquet format.
Every row is timestamped with a \textit{serializer sequence number}
(SSN)), which is an identifier for each transaction in the serial order. 
The SSN is used by queries to 
antijoin with \texttt{ConstraintFailures} table 
to remove entries from transactions that failed aggregation constraints (checking constraints
 happens after publish).

$\bullet$ Removing deltas from conflict rollbacks: \textsc{Publisher} also performs 
an antijoin with \texttt{Rollbacks} table to remove all rows from transactions that hit 
read-write conflicts.

$\bullet$ Updating indexes: OLTP queries often do highly selective (``point'') lookups
that need indexes to be efficient, so \textsc{Publisher} has to update these as well. 
\wiser partitions the primary key index the same way as the data, so that update is also a local operation.
We use an LSM-based indexing structure~\cite{o1996log} so that incremental inserts are efficient and incur low latency even
on large tables.
Maintaining secondary indexes in \wiser is a more
expensive, shuffle-like data movement, so we expect to maintain only a limited number of secondary indexes at \textsc{Publish} stage.
\wiser implements the HERMIT technique~\cite{wu2019designing,wu2019hermit} to reduce the storage consumption caused by secondary indexes.

Experimentally we find \textsc{Publish} stage to be the throughput bottleneck for \wiser, and
so \textsc{Publish} is done in parallel, across all nodes. This means publication into
tables can happen \textit{out of serial order}. So we maintain a
\texttt{PublishFrontiers} table, whose schema is identical to
\texttt{SerializeFrontiers}.  Each row of this table contains a node ID and an LSN
up to which log entries from that node have been published.  Queries access
published tables (and primary indexes) to access all content up to the \textit{minimum} 
publication frontier (across all nodes). Beyond that (until the
serialization frontier), queries need to scan logs.  Most transactions avoid
this scan by waiting until the publication frontier catches up to their
snapshot, which they picked up by reading \texttt{SerializeFrontiers} at
transaction start. In this way, they only need to scan logs if they have very tight latency
requirements.

One thing that \textsc{Publish} does \textit{not} do is table repartitioning: all the data ingested 
at one node stays at that node. Clients can carefully route transactions
to respect any desired clustering properties. For example, they may send a transaction 
that inserts a fact row to a node based upon the hash of the primary key or based on geography. 
\textsc{Publish} simply converts logs to table format, local to each node.

\subsection{Node Departures and Re-Joins}
\label{joinsleaves}

\wiser has a \texttt{Nodes} table that contains the list of nodes.
Thus node join events get serialized 
just like other database modifications. 

Node departures from the network are not supported directly. If a 
node becomes unresponsive, its peers cannot directly evict that node, because 
all the transactions that reached Promise on that node must still be serialized.
This is handled by the node replicas.
When the node re-joins the cluster, it gets a new node ID.

\section{Conflict Resolution}
\label{sec:conflicts}

Conflicts in \wiser are resolved after we have determined a serial order.
Thus the resolution problem is simple: 
the reads in a transaction $X$ saw a snapshot that is earlier than
where its delta fits in the serial order. $X$'s delta includes $X$'s read-set and $X$'s snapshot 
(seen SSN).
So the \textsc{ConflictResolver} needs to traverse the 
database chain of deltas, starting from the prior resolved point, and verify
for each delta whether its read-set could have changed in value due to intervening deltas
(between $X$'s snapshot and $X$). If it could have changed, then $X$ is added to \texttt{Rollbacks} table.
$X$ is stored as a transaction ID in \texttt{Rollbacks} table: SSNs are assigned later, at \texttt{Publish},
only for the successful transactions.

Note that this process can give false positives -- excess rollbacks, for two reasons.
First, read-sets and write-sets
are tracked approximately as we describe below. Second, in our data structure it is expensive to remove 
the writes of a rolled-back transaction, so a transaction can roll back due to conflicts with another one, 
which rolled-back 
for other reasons. This behavior is similar to the way deadlocks are typically handled, and applications 
are used to retrying transactions upon such roll-backs.

\subsection{Representation of Read Sets}

\wiser represents read-sets as 
entries in a \texttt{ReadWriteSet} table. 
This ``all state is in tables'' principle means 
that a transaction's delta naturally incorporates its read-set as well.
Queries are allowed to selectively read data by specifying conjunctive
equality predicates on one or more columns, or range predicates 
(for table scan queries, the read set is escalated to the full table).
For the common case of equality predicates, 
we compute a single 64-bit hash value over the key,
thus easily handling different data types and multi-column predicates.
For range predicates we store the boundaries of the range.

Our current implementation of \wiser unifies the role of \textsc{ConflictResolver} and \textsc{Serializer}. Thus
each node running transactions replicates \texttt{ReadWriteSet} to the \textsc{Serializer} node,  
along with 
its regular heartbeats (the one containing 
LSN ranges of recent additions to its log).

\subsection{Representation of Writes}

To resolve conflicts, the read-set of a transaction $X$ needs to be compared against
all writes in the serialization order between $X$'s snapshot and $X$. Recall
that \textsc{Publish} timestamps every row with its SSN. Thus this comparison
can be done by an index lookup when available, and if not then via a 
table scan over recent data (\wiser maintains a min-max synopsis on each block
for each column, so this is not terribly inefficient).

Still, to get greater efficiency in conflict resolution,
the \textsc{Serializer} caches recent writes in memory directly.
When a transaction $X$ is during execution, the read-set of $X$ is placed in \texttt{ReadWriteSet} as
discussed earlier. 
In addition, the primary keys of rows written by $X$ 
are also placed in \texttt{ReadWriteSet}.
Notice that the reads are represented logically (e.g., as key ranges), whereas writes are enumerated row-by-row.
As the \textsc{Serializer} receives writes, it maintains a list of hash tables.
These hash tables map hash(key) onto the \textit{latest} SSN when that key was modified 
(using the same function to map arbitrary composite key values
onto a 64-bit hash as we did for the reads).

The \textsc{Serializer} streams all read-sets it receives, looking for conflicts in the recent writes.
A read conflicts with a write only if (a) the key-range of the read spans the
key of the write, and (b) the SSN of the write is greater than
the snapshot of $X$.

These writes are maintained as a linked list of hash tables to facilitate eviction. 
Each hash table holds writes within a range of SSNs.
So periodically we drop hash tables for SSN ranges that are older than
most transactions.

\subsection{Catching Phantoms from Range Predicates}

Queries with range predicates are challenging to run serializably because of phantoms.
The read-set for such a query is a range of key values, but the DBMS must guard against concurrent 
inserts of any value in that range. Typically this is done by placing
a ``next-key'' lock on the
index tree structure -- i.e., a guard on the values just beyond the two boundary values of that
range~\cite{mohan1989aries}.
The trouble with this approach is that it 
is efficient only when the index is memory-resident.
The space taken to check for conflicts is proportional to the database size.

\wiser handles conflicts due to range predicates very similarly to 
the case of equality predicates. When there is a range index defined on a column,
recent writes are cached at the \textsc{Serializer} as a in-memory tree of column values (not hashes).
This is also a list of trees, organized by ranges of SSNs, much like the hash tables.
Conflict detection is done by probing read-sets into these trees, and checking
any matching writes for SSNs beyond the snapshot of the read-set.

\section{Constraint Resolution}
\label{sec:constraints}

We have now built up enough machinery to discuss how aggregation constraints
are checked. These constraints are checked \textit{after} conflict resolution and
\textit{after} publish, who has stamped each row with the SSN (its position in the
serial order).  

Logically, constraint resolution is a walk through the published data in SSN order,
from the prior resolved point all the way up to the latest published point. The
\textsc{ConstraintResolver} maintains a running aggregate, and we only support
incrementally computable, algebraic aggregation constraints. 
The constraint is specified as a predicate on the result of the aggregation. In the online shopping application,
the aggregation is to compute the sum of quantities grouped by product IDs, and the predicate is to check whether the aggregated 
result is larger than or equal to 0.

The \textsc{ConstraintResolver} can run queries
repeatedly to update the running aggregate, each time selecting data in the
range from the prior resolved SSN to a value $t$, 
where $t$ is gradually increased until it hits a
constraint violation.

But every transaction creates a new SSN, so this approach is far too slow.
We want constraints to be evaluated on batches of transactions, corresponding to
the batches that were serialized from each node. The challenge is that aggregates are generally
not monotonic: just because aggregation constraints are not violated after incorporating
an entire batch of transactions, we cannot assert that the constraints were not violated 
at some intermediate point (for example, a bunch of customer orders might have depleted inventory
to be negative, and then a restocking transaction might have been serialized).

A second design goal is to do parallel constraint resolution, in two different forms:

$\bullet$ Source parallelism: We want the bulk of aggregation to be performed
at the nodes running the OTLP transactions, that reduces data movement and also gives
significant parallelism.

$\bullet$ Resolver parallelism: The running aggregation state for constraint enforcement 
can be large: for example, it is the running inventory of each product in a retail application. So we want 
individual resolvers to focus on subsets of products.

\subsection{Partial Constraint Evaluation}

\wiser supports partial constraint evaluation to reduce 
computation complexity.

Let us consider a setting with three nodes, each with 
multiple batches of transactions.
The \textsc{ConstraintResolver} repeatedly picks up one batch from each node to resolve
(recall that \texttt{SerializeFrontiers} lists the batches).
It first pushes down to each node a partial aggregation query, which is derived from the constraint
(e.g., compute the sum of quantities grouped by product IDs, in the online shopping application). 

Next, the \textsc{ConstraintResolver} uses the partial aggregation results from each node
to compute prefix sums (per the serial order) 
for the value of the
aggregate at the start of each batch. If the prefix sum after any batch
violates the constraint, the \textsc{ConstraintResolver} reduces that batch's contribution to be one 
that makes the constraint just satisfied. For example, if the prefix sum before and after a batch is
12 and -5, respectively,
then the \textsc{ConstraintResolver} artificially inflates 
the prefix sum after that batch to 0. 
This is not to abort the whole batch; instead,  
it is only to avoid 
polluting the prefix sums for other batches. 
In effect, the \textsc{ConstraintResolver} is delegating
to the nodes the responsibility of keeping the constraint satisfied by rolling back any 
transactions that cause violations.

After that, the \textsc{ConstraintResolver} communicates these prefix sums to each node, and asks it 
to re-evaluate the aggregation query, except this time with a modified aggregation function.
This new query runs the same group-by, but initializes its 
aggregation hash table with the prefix sum from the \textsc{ConstraintResolver}. 
The aggregation scans the input in SSN order (this does not involve an extra sort,
because the rows were published in that order). As the query maintains the running aggregate
it continually applies the constraint (e.g., inventories should stay positive),
and surgically identifies and excludes individual transactions that cause constraint 
violation.

As a side-effect of this query, each node now adds rows to the \texttt{ConstraintFailures} 
table -- these rows contain just the SSN of the constraint-violating transactions.

Notice that excluding these transactions will cause the prefix sums to change, but this 
is \textit{not communicated} back to the \textsc{ConstraintResolver}. After all, removing transactions
that caused constraints to be violated only moves the running aggregate in a beneficial 
direction, so it is safe for other nodes (checking other batches)
to ignore this effect.

Constraint resolution is the last stage of transaction processing. Queries do 
antijoins with \texttt{ConstraintFailures} to avoid seeing changes from 
transactions that fail constraint resolution. Transactions
that reached \textsc{Promise} but failed conflict resolution or constraint 
resolution are aborted transactions.

\section{Performance  Evaluation}
\label{expts}

\wiser is the next version of the Wildfire 
research prototype~\cite{barber2016wildfire}.
It is a HTAP DBMS that currently supports ingest, via Apache Spark and via a custom API; queries, directly
via a SQL dialect, as well 
as ones pushed-down from Spark, and now multi-statement transactions. The last feature 
is the focus of this section.
Transactional changes are replicated and hardened to a quorum of nodes.
Tables can optionally have a primary key index on one or more columns, and this index supports a mix of equality
and range predicates.
All tables and logs are organized as Apache Parquet blocks. We use the \texttt{parquet-cpp} writer.

All experiments in this section
are run in a Docker environment, on 
a machine cluster with 28 nodes and 1.6 TB DRAMs in total. 
Writes are hardened to PCI-e attached SSDs 
with \texttt{fsync}. We execute one transaction thread in each 
node; the background operations such as serialization, conflict resolution, and
publication, are all implemented on separate nodes from those running transactions.

\subsection{Throughput with No Conflicts}

Our first experiment investigates the basic transaction processing throughput in \wiser, 
without considering conflicts or rollbacks. 
Transactions are run up to \textsc{Serialize}.  
\textsc{ConflictResolver} does nothing, because there are no conflicts. 
We do not run \textsc{Publish} for this experiment.

Our workload consists of just the \texttt{NewOrder} transaction from 
the online shopping application discussed in \cref{introduction}, 
run repeatedly, with no think-time,
and no aggregation constraints.
\texttt{Products} table has 10,000 items, with an index on \texttt{productId} column. Each transaction 
looks up the current price of 10 products 
and inserts an order containing those 10 products into \texttt{Orders} 
table.
The values in \texttt{OrderId} column are generated as sequential numbers, and there are no price updates, 
so in this experiment there are no conflicts.

\begin{figure}[t!]
  \centering
  \includegraphics[width=0.9\columnwidth,clip,trim={2 2 2 2},]
    {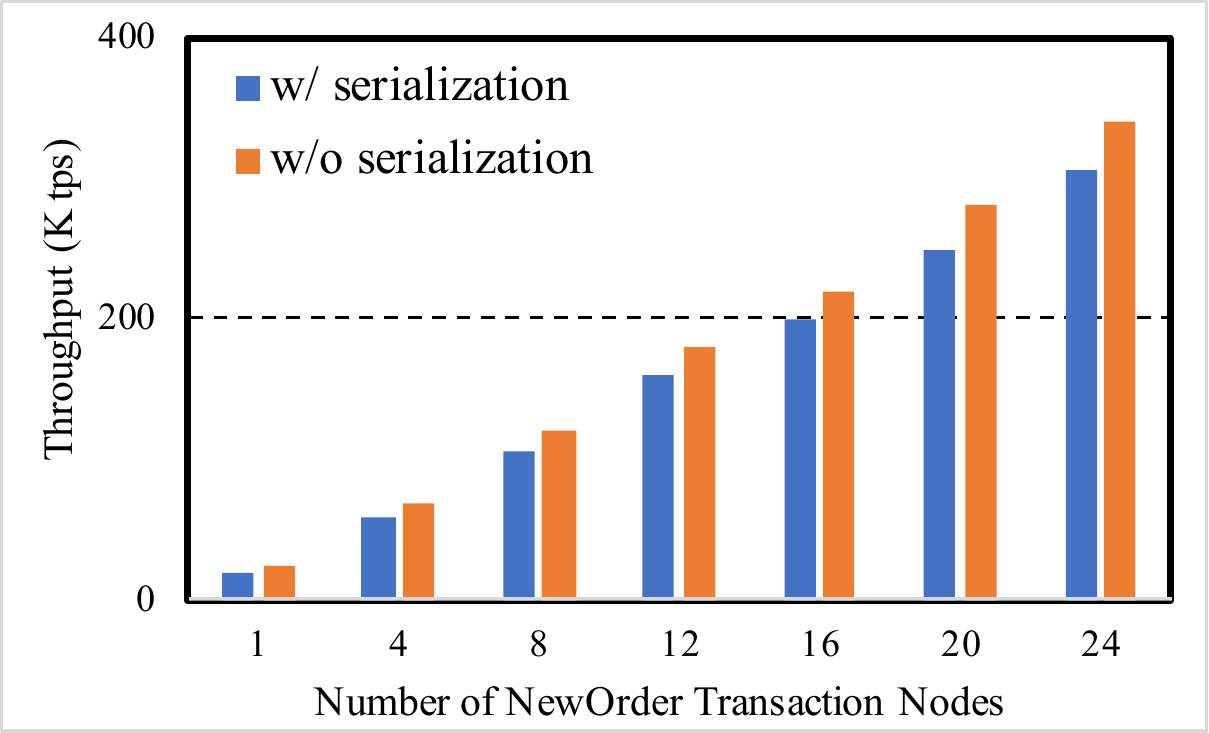}
  \caption{\texttt{NewOrder} transaction throughput 
  as a function of nodes, 
  with and without serialization.}
  \label{figure:exp1}
\end{figure}

\cref{figure:exp1} shows the transaction throughput we obtain 
with this simple workload, and the scaling. 
Notice that at 24 nodes we are inserting 3.5
million rows/s into \texttt{Orders} table, 
and correspondingly doing 3.5 million lookups/s into the
index on \texttt{Products} table. 
In this setting, the X-axis is the number of nodes running the 
\texttt{NewOrder} transaction.
We have a separate node running the \textsc{Serializer} and \textsc{ConflictResolver}.
Notice also that we scale well; our system has pretty much no contention,
except for the log writes (which is implemented as a lock-free linked list of blocks).
The same plot also shows what happens if we turn off 
serialization: the effect is negligible.

Note that each transaction implicitly also generates 10 reads and 10 writes, which are 
tracked and sent to the \textsc{Serializer}.
From profiling we verify that most of the time goes into this
tracking of read-write sets.

\subsection{Throughput with Conflicts}

Next, we add conflicts, by introducing price updates for the products.
We run a \texttt{UpdatePrice} transaction every 100 ms, 
and that updates the price for a certain number
of randomly chosen products.
This introduces read-write conflicts because the price that a 
\texttt{NewOrder} transaction looks up might become stale by 
the time it goes to \textsc{Serialize} stage. 

\begin{figure}[t!]
    \centering
    \includegraphics[width=0.9\columnwidth,clip,trim={2 2 2 2},]
      {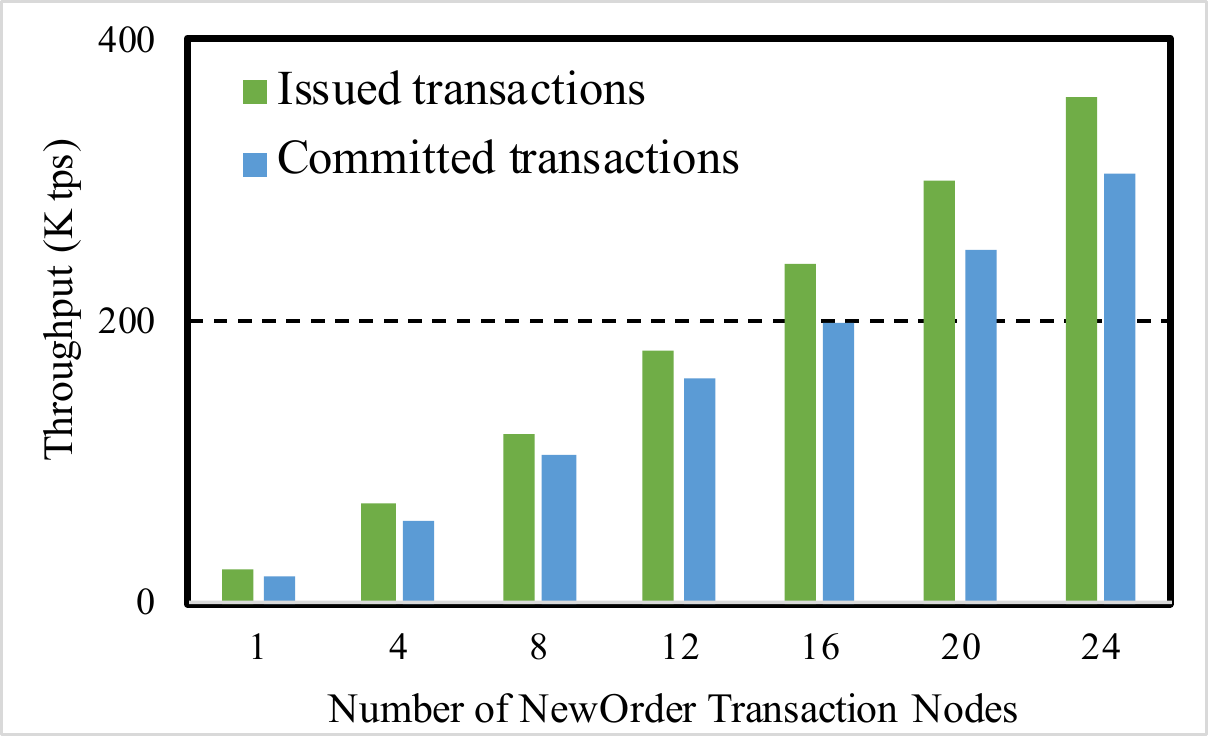}
    \caption{\texttt{NewOrder} transaction throughput 
    as a function of nodes, with \texttt{UpdatePrice} transaction size
    set to 10.}
    \label{figure:neworder_updateprice_serresolve10}
\end{figure}

\cref{figure:neworder_updateprice_serresolve10} and
\cref{figure:neworder_updateprice_serresolve10} show the \texttt{NewOrder} transaction throughput, for both issued and committed transactions,
as a function of the 
number of nodes executing \texttt{NewOrder} transactions. Recall that 
there is a separate node for serialization and conflict resolution,
and the \texttt{UpdatePrice} transactions are issued in another thread.
We plot this at two price update volumes: where we update 10 prices 
in each \texttt{UpdatePrice} transaction,
and where we update 20 prices in each \texttt{UpdatePrice} transaction. 

Notice that the issued and committed throughputs are almost identical. 
We measure about 1\% rollback rate 
when we update 10 prices at a time, and 2\% when we update 20 prices at a time.

There are 10,000 products,
and each transaction is looking up 10 random products. Our \textsc{Serializer} picks up batches of transactions to 
serialize
roughly every 100 ms. Thus, if the \textsc{Serializer} keeps up perfectly with the transaction rate,
each \texttt{NewOrder} transaction will overlap with about one \texttt{UpdatePrice} transaction (in the sense that the snapshot that
the \texttt{NewOrder} sees will be behind its serialization point by about one \texttt{UpdatePrice} transaction).

This expected behavior matches the rollback rate well.
When an \texttt{UpdatePrice} updates 10 prices,
a given product has a $10/10000$ or 0.001 chance of getting updated. So the chance that 
a transaction will commit is $(1-0.001)^{10}$, or about 0.99.
Likewise, when an \texttt{UpdatePrice} updates 20 prices, 
the chance that a transaction will commit is
$(1-0.002)^{10}$, or about 0.98, which matches our observed rollback rates.

\begin{figure}[t!]
    \centering
    \includegraphics[width=0.9\columnwidth,clip,trim={2 2 2 2},]
      {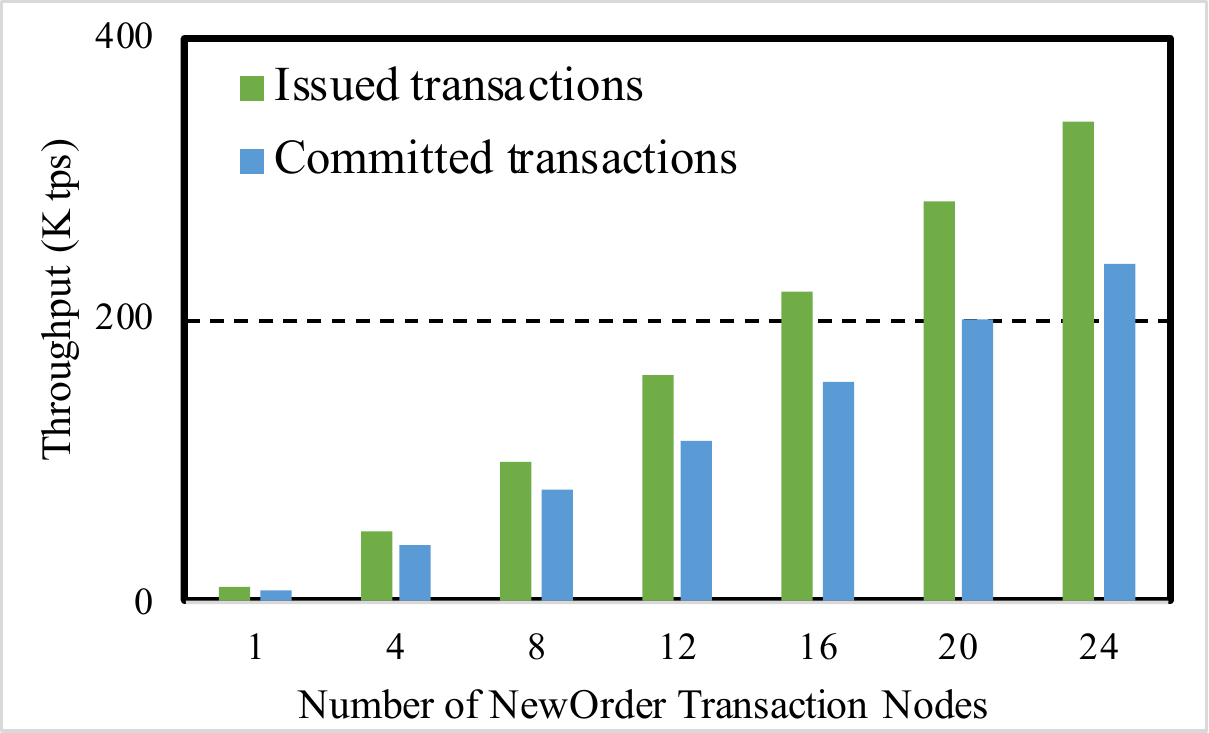}
    \caption{\texttt{NewOrder} transaction throughput 
    as a function of nodes, with \texttt{UpdatePrice} transaction size
    set to 20.}
    \label{figure:neworder_updateprice_serresolve20}
\end{figure}

\subsection{End-to-End Throughput}

\begin{figure}[t!]
    \centering
    \includegraphics[width=0.9\columnwidth,clip,trim={2 2 2 2},]{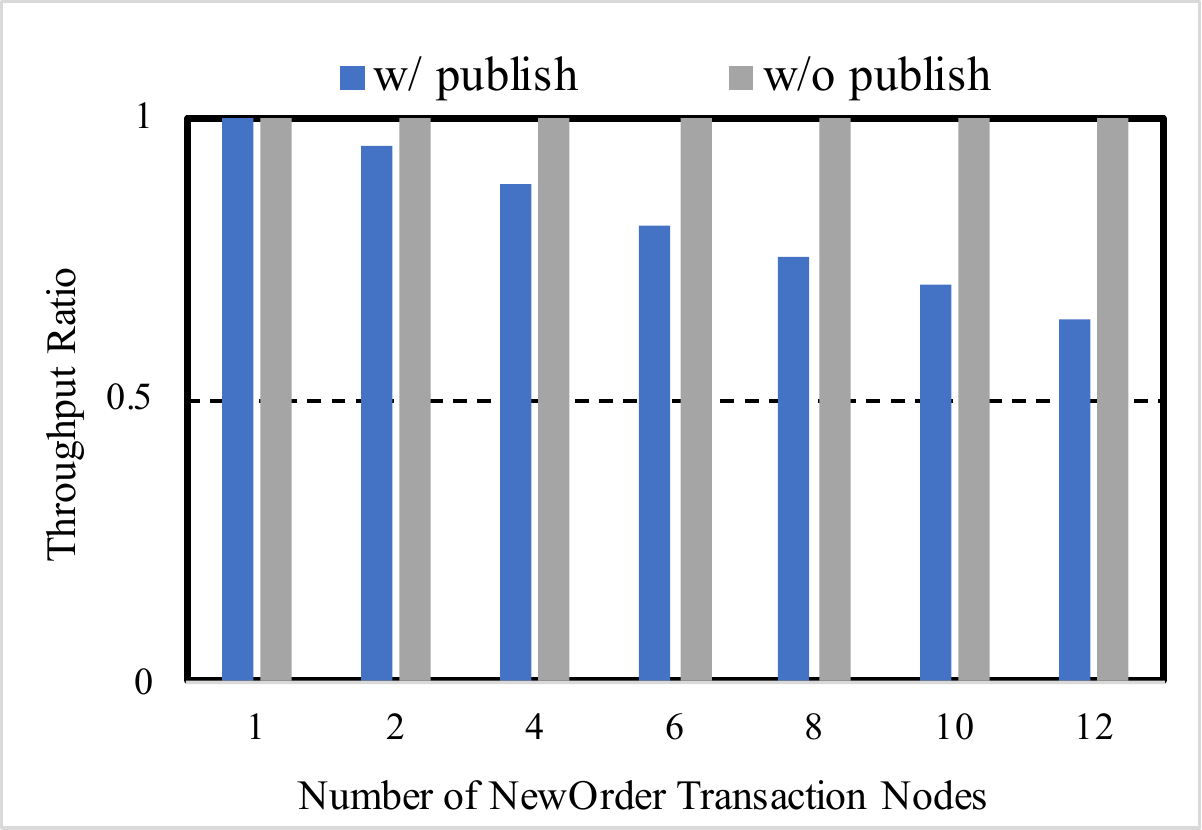}
  \caption{\texttt{NewOrder} transaction throughput ratio 
  as a function of nodes, 
  with and without publish.}
    \label{figure:publish}
\end{figure}

Our next experiment looks at the end-to-end transaction throughput. 
We add the \textsc{Publisher},
and hence what we measure is the throughput of published transactions.
We have one \textsc{Publisher} per (logical) node, so we need as many \textsc{Publisher} nodes
as \texttt{NewOrder} nodes. Thus this experiment scales only up to 12 nodes.
(12 nodes for \texttt{NewOrder}, one for \texttt{UpdatePrice}, one for \textsc{Serializer} and \textsc{ConflictResolver}, 12 for \textsc{Publisher}).
\cref{figure:publish} shows the effect of adding publish on the 
system throughput. There is about a 70\% hit on throughput, and our scaling is affected. \textsc{Publisher}
has to write data to Parquet blocks, and we find this to be a bottleneck. 

\subsection{Latency Analysis}

\begin{figure}[t!]
    \centering
    \includegraphics[width=0.9\columnwidth,clip,trim={2 2 2 2},]{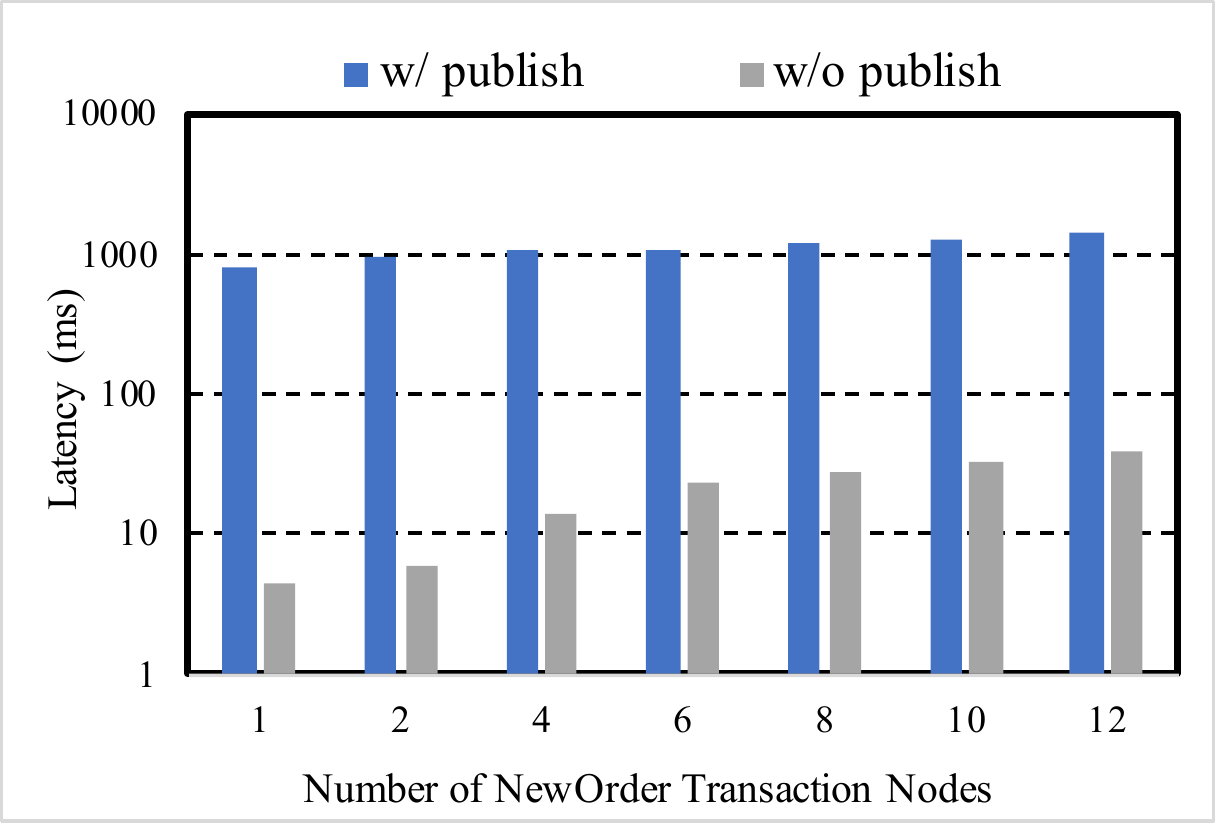}
  \caption{\texttt{NewOrder} transaction latency as a function of nodes, 
  with and without publish.}
    \label{fig:latency}
\end{figure}

Our final experiment looks at transaction latency.
We rerun the full experiment, with \texttt{NewOrder} and \texttt{UpdatePrice} transactions,
as well as \textsc{Serializer} and \textsc{ConflictResolver}, 
and \textsc{Publisher}.
This time we measure the average \texttt{NewOrder} transaction
latency, until \textsc{Publish} and \textsc{Serialize} stages.
\cref{fig:latency} plots this latency as a function of the number of \texttt{NewOrder} threads.
Observe that latency to serialize is in the tens of ms, and 
this is the minimum latency until a query can access the transaction's changes
(by reading the log).
Going to publish pushes the latency to about one second, this is due to overheads
in forming Parquet blocks.

\section{Related Work}
\label{sec:related-work}



 

%
%
%

\wiser builds on and borrows hugely from the large volume of literature on transaction
processing, where both shared-nothing and shared-storage models have been heavily studied~\cite{rstar,pdb}.

After many efforts on NoSQL databases that downplayed the importance of consistency,
the last decade has seen a renewed interest in systems
that provide strongly-consistent transactions over large-scale distributed systems,
across partitions and replicas.
One type of distributed DBMS, represented by IBM's
Spinnaker~\cite{rao2011using} and Google's Spanner~\cite{corbett2013spanner},
adopts 2PC to coordinate distributed transactions and exploits consensus protocols
like Paxos~\cite{lamport1998part} or Raft~\cite{ongaro2014search} to
synchronize replicas during transaction execution. 
In Spanner, write transactions consult a lock server, resembling classical shared-nothing DBMSs.

Another approach, taken by systems like Calvin~\cite{thomson2012calvin} and FaunaDB~\cite{faunadb}, is to 
use a deterministic execution model. These DBMSs coordinate transactions within a pre-processing phase. 
In this phase, the DBMS must extract the write-set from the to-be-executed transaction, and accordingly
choose a serial order. \wiser, like Calvin, tries to come up with a serial order,
 and then efficiently do 
post-processing on the deterministic state. But unlike Calvin, \wiser does not pre-process transactions
and thus can support transactions that are not pre-analyzable stored procedures.

Many modern DBMSs attempt to do analytics with transactions~\cite{ozcan2017hybrid}.
HyPer~\cite{kemper2011hyper} is a modern main-memory DBMS that targets efficiently 
supporting OLTP and OLAP workloads in the same memory space. It leverages multi-version 
concurrency control (MVCC) and precision locks to achieve fully serializable transaction
processing without blocking on-the-fly analytical queries.
Similar to HyPer, Peloton~\cite{pavlo2017self} also leverages MVCC~\cite{wu2017empirical} to
isolate on-line transactions from analytical queries, but it also leverages hybrid 
storage layout to better support different types of queries~\cite{arulraj2016bridging}.
Focusing on modern hardware, Appuswamy et al.~\cite{appuswamy2017case} studied the case
for developing HTAP DBMSs by exploiting the power of both CPUs and GPUs. 
A number of commercial systems also claim HTAP features, such as 
SAP HANA~\cite{hana} and IBM DB2 Event Store.
SQL Server introduced their enhancements for hybrid workloads in 2016~\cite{larson2015real}.
By improving support for column-store indexes, SQL Server can 
enable real-time analytics concurrently with transactional processing.

DBMSs have long leveraged sharding techniques to store data across multiple nodes. 
Sharding simplifies scaling, but does 
bring in distributed transactions, which can significantly degrade system scalability.
Harding et al.~\cite{harding2017evaluation} performed an experimental evaluation of 
distributed concurrency control protocols.
Some systems use transaction chopping to serialize transactions~\cite{zhang2013transaction,mu2014extracting,wu2017fast}.
H-Store/VoltDB~\cite{kallman2008h,stonebraker2013voltdb} uses partition-level locks
to reduce the overhead of distributed operations.
Other systems try to avoid data partitioning.
For example, Hyder~\cite{bernstein2011hyder} exploits SSDs to scale DBMSs 
in a shared-flash setting, whereas FaRM~\cite{dragojevic2015no} uses RDMA and non-volatile 
DRAM to avoid user-level partitioning.

\section{Conclusions and Future Work}
\label{sec:conclusion}

The distributed systems community has developed consensus protocols
mostly independently from the database community's struggles with distributed transactions.
We have described one approach to exploiting consensus as the underlying mechanism for 
transaction serialization, and shown it provides two key benefits: higher availability
during the application's interaction with the running transaction (the promise step),
and greater scaling and easier enforcement of aggregation constraints after serialization.

\wiser comes at a time when demand for HTAP and real-time analytics is exploding. Yet, 
OLTP systems are complex and well-entrenched ``mission-critical'' pieces  of applications.
So we have an uphill task to penetrate this market, requiring that we massage our 'promise-serialize' 
step to fit with existing applications, equal or surpass the plethora of must-have OLTP features 
(meeting existing isolation semantics, handling referential integrity, etc.),
and demonstrate value from pulling compensation logic out of applications.

\wiser's design of putting all state into tables, and having a tiny consensus piece,
may also help in Byzantine settings. Blockchain systems are to some extent bypassing and 
re-inventing
many aspects of classical OLTP. We want to explore whether casting OLTP as a consensus
problem will make it easier to provide higher security, timestamping, and non-perturbation 
guarantees within existing DBMSs.


\bibliographystyle{abbrv}
\small
\bibliography{acid}


\end{document}